\documentclass[twocolumn,showpacs,preprintnumbers,amsmath,amssymb]{revtex4}
\usepackage{graphicx}
\usepackage{dcolumn}
\usepackage{bm}
\usepackage{here}
\usepackage{tabularx}
\begin{document}
\title{Boolean Dynamics of Kauffman Models with a Scale-Free Network}
\author{Kazumoto Iguchi}
\email{kazumoto@stannet.ne.jp}
 \affiliation{70-3 Shinhari, Hari, Anan, Tokushima 774-0003, 
Japan}
\author{Shuichi Kinoshita} 
\email{f01j006g@mail.cc.niigata-u.ac.jp}
 \affiliation{Graduate School of Science and Technology, Niigata University,
Ikarashi 2-Nochou 8050, Niigata 950-2181, Japan}
\author{Hiroaki S. Yamada}
\email{hyamada@uranus.dti.ne.jp}
 \affiliation{
 Yamada Physics Research Laboratory, 5-7-14 Aoyama, Niigata 950-2002, Japan
 }%
\date{\today}

\begin{abstract}
We study the Boolean dynamics of the "quenched" Kauffman models with a directed scale-free network,
 comparing with that of the original directed random Kauffman networks and that of the directed exponential-fluctuation networks.
We have numerically investigated the distributions of the state cycle lengths and its changes 
as the network size $N$ and the average degree $\langle k \rangle $ of nodes increase.
In the relatively small network ($N \sim 150$),
the median, the mean value and the standard deviation grow exponentially
with $N$ in the directed scale-free and the directed exponential-fluctuation networks with $\langle k \rangle =2 $, 
where the function forms of the distributions are given as an almost exponential. We have found that for the relatively large 
$N \sim 10^3$  the growth of the median of the distribution over the attractor lengths asymptotically  changes from algebraic type 
to exponential one as the average degree $\langle k \rangle $ goes to $\langle k \rangle =2$.
The result supports an existence of the transition at  $\langle k \rangle_c =2$ derived in the annealed model.
\end{abstract}
\pacs{89.75.Hc,05.40.-a}
\maketitle

\section{Introduction}
The origin of life has attracted many scientists as one of the unsolved problems in science for a long time\cite{Maddox}.
To answer the quest, the self-organization of matter\cite{Eigen}
and the emergence of order\cite{KauffmanBook1}
have been regarded as the key ideas.
Kauffman first introduced the so-called {\it Kauffman model} --
a random Boolean network (RBN) model,
based upon the random network
theory\cite{ErdosRenyi1,ErdosRenyi2,ErdosRenyi3}.
This model has become a prototype for many authors to study complex systems
 such as metabolic stability and epigenesis, genetic regulatory networks, and
 transcriptional networks
 \cite{KauffmanBook1,Kauffman1,Kauffman2,SawKauff,Kauffman3,Kauffman4,SocKauff,
 KauffPST}as 
 well as general Boolean networks
 \cite{DerridaPomeau,DerridaWeis,DerridaStauf,FlyvLaut,Altenberg,AldanaCK},
 neural networks\cite{Hopfield} and
 spin glasses\cite{Anderson,SteinAnder,Anderson2}.
In the RBN, we assume that the total number of the elements (i.e., nodes) $N$
 and the link number of an element (i.e., the degree of a node) $K$
 in a directed random network are kept to be constants, respectively.
 Kauffman found numerically that there is a phase transition
 from an {\it ordered} phase to a {\it chaotic} phase through the
 {\it critical} point (i.e., edge of chaos) at $K = K_{c} = 2$
 for the equally probable model of the Boolean functions for 0 and 1.
 Here in the chaotic phase {\it the number of the state cycle attractors}
 grows as exponential in $N$;
 in the ordered phase the growth is proportional to $N$;
 in the critical point the growth is proportional to $\sqrt{N}$.
 Later, this phenomenon has been analytically verified by studying
 the "annealed" Kauffman
 models\cite{DerridaPomeau,DerridaWeis,DerridaStauf}.
 And also exact results have been analytically obtained for the special cases
 of $K =  N$\cite{DerridaFlyv1,DerridaFlyv2,DerridaFlyv3,DerridaFlyv4,Derrida,Flyvbjerg} and $K = 1$
\cite{FlyvbjergKjaer,SamuelTroein05,DrosselMG}.
Recently there has appeared a revival interest on the study of the critical
 point of the
 Kauffman models with $K_{c}=2$
 as well as $K_{c} = 1/\rho$ where $\rho = 2p(1-p)$
 and $p$ denotes a probability ($0 \le p \le 1$) given in the next section 
\cite{BhattaLiang,BasParisi1,BasParisi2,BasParisi3,BasParisi4,AlbertBarabasi,SocKauffman,SamuelTroein03,KlemmBornholdt,CKZ1,CKZ2}.
Here, the power-law distributions of the cycle length of
attractor\cite{BhattaLiang}, the models with the finite numbers of $1 < K < 4$\cite{BasParisi2,BasParisi3,BasParisi4}, 
the special cases of $K_{c} = 2$ networks such as the one-dimensional network and the Cayley tree\cite{AlbertBarabasi},
the scaling nature of the critical ($K_{c} = 2$) point and the chaotic ($K > 2$) phase\cite{SocKauffman}, 
the superpolynomial growth of the number of the state cycle attractors\cite{SamuelTroein03}, 
the stability of the system\cite{KlemmBornholdt} have been studied both numerically and analytically.
 These results suggest that the critical point of Kauffman model with $K_{c} = 2$ is very special.
Especially, Bastolla and Parisi\cite{BasParisi2,BasParisi3,BasParisi4}
elucidated
that what is important on the critical line is the effective connectivity
$K_{eff}$ ($\approx 1$) and that
the critical point of the model exhibits the nature of
percolation of information
where the character shares much with that of $K = 1$ studied
by Flyvbjerg and Kj\ae r\cite{FlyvbjergKjaer}.
To study this rather peculiar aspect further,
Coppersmith et. al.\cite{CKZ1,CKZ2} studied numerically
the "reversible" Kauffman model that is
different from the original one in the sense that
the former is dissipative while the latter is non-dissipative.
They supplied a critical value of $K_{c} \cong 1.62$ for the non-dissipative
cases.
On the other hand, at nearly the end of 1990's
 the scale-free networks have been discovered from studying the growth of the
 internet geometry and topology
 \cite{Strogatz,Barabasi,AlBarabasi1,BaraBona,Newman}.
 After the discovery, scientists have become aware that many systems such as
 those which were originally studied by Kauffman
 as well as other various systems such as internet topology, human sexual
 relationship, scientific collaboration, economical network, and so on.
 belong to the category of the scale-free networks
 \cite{Strogatz,Barabasi,AlBarabasi1,BaraBona,Newman}.
 Therefore, it is very natural to apply the concepts of scale-free networks
 to the Kauffman model and to ask whether or not there appears the difference
 between the RBN and the scale-free random Boolean network (SFRBN).
Recently, as to this direction, the dynamics of the "quenched" SFRBN
 has been intensively studied
\cite{LuqueSole,AlBarabasi2,FoxHill,OosawaSava,Wang,Aldana1,AldanaCl,Aldana2,SerraVA,CastroSM,Iguchi05}.
It has been shown that the critical point occurs
when the average degree $\langle k\rangle$
of the scale-free networks equals $\langle k\rangle_{c} = 2$
for the equal probability (i.e. no bias) models of $p = 0.5$,
analytically studying the "annealed" SFRBN
\cite{LuqueSole,Aldana1,AldanaCl,Aldana2}
that are the scale-free analogs of the "annealed" RBN
 studied by Derrida and Pomeau\cite{DerridaPomeau}.
Analytical results have been obtained for the annealed model with $N \to
\infty$
under some assumptions such as mean field approximation and ergodicity.
On the other hand, a lot of numerical results have been calculated for
finite-size quenched model.
In the numerical simulation there are various ways for the sampling over,
for
example, (A)different networks, (B)different initial states, and
(C)different Boolean functions assigned for each node.
In particular, the mean value over the numerical distribution becomes
insignificance
 when the distribution is broad because the order of fluctuation exceeds
over
the mean value. Then the relative fluctuation diverges for
system size going to infinity, i.e. not self-averaging.
Such a phenomena can be found at second-order phase transition
and scale-free networks with power-law correlation \cite{stauffer04}.
In the situation
the median value is statistically more robust against
the artifacts due to undersampling of the set of the networks and the
initial states than the mean value.
In this paper
we would like to mainly investigate a qualitative transition in the function
form of the median value $\bar{m}$ over the distribution of attractor length
as a function of the {\it finite $N$}, changing the average degree $\langle
k \rangle$.
Note that the transition does not always mean phase transition in the usual
statistical mechanics sense, which is defined in the thermodynamic limit of
$N \to \infty$.
One of our interests is also finite-size effect
 of the transition phenomena around $\langle k \rangle =2$ in the Boolean dynamics.
{\it How is the transition observed
in the relatively small network
corresponding to the realistic network size?}

To ease the reference, we first summarize our main result:
There is a transition of the function form $\bar{m} \propto N^\alpha$ from
$\alpha <1$
 to  $\alpha > 1$ within a range of $1.4 < \langle k \rangle_{c} < 1.7$
in the "quenched" SFRBN without bias.
 The transition becomes important when we consider some realistic biological
networks,
as Kauffman first introduced, because the realistic data suggest that the
number of cell types
in organism is crudely proportional to the linear or square-root function of
the DNA content
per cell, $\propto N^\alpha (\alpha \leq 1)$.
We adapted a {\it completely synchronous updating rule} in the Boolean
dynamics
because the attractor lengths are well-defined only for the synchronous
updating rule.
However, it is noting that the synchronism idealization is not always true
for biological systems as genetic regulatory networks and asynchronous
updating rules are more plausible for biological phenomena\cite{harvey97}.
Moreover, the function form $\bar{m}(N)$
asymptotically changes from the algebraic type  $\bar{m}(N) \propto
N^\alpha$ to the exponential one 
as the average degree $\langle k \rangle $ goes to $\langle k \rangle =2$.
This result is consistent with the previous belief that
the transition occurs at the critical value of the degree of nodes of $K_c =
2$
in the "annealed" RBN and SFRBN s\cite{KauffmanBook1}.
 The organization of the paper is the following.
 In Sec.II, we present the Kauffman models that we study.
 In Sec.III, to apply various networks to the Kauffman models
 we give how to generate such networks.
 In Sec.IV, we study the cycle distributions of the "quenched" Kauffman models for the various network systems.
 In Sec.V, conclusions will be made.
 We mainly focused on the change of the functional form of the median
 around the critical value $\langle k \rangle \leq 2$ in the main text.
 However, the other quantities such as Derrida plots and frozen nodes density
 are also used in order to investigate the Boolean dynamics.
 We briefly give analytical result for $\langle k \rangle =1$ in appendix A
 and Derrida plots in appendix B.
%

%
\section{RANDOM BOOLEAN NETWORK (RBN)}
The RBN requires us to assume that
the total number of nodes (vertices) $N$ and
the degree (the number of links) of the $i$-th node $k_{i}$ are fixed in the
problem,
where all $k_{i} = K$.
Since there are $K$ inputs to each node,
$2^{2^{K}}$ Boolean functions can be defined on each node;
the number $2^{2^{K}}$ certainly becomes very large as $K$ becomes a large
number.
Then, we assume that Boolean functions are randomly chosen on each node
from the $2^{2^{K}}$ possibilities.
Locally this can be given by $$\sigma_{i}(t+1) = B_{i}(\sigma_{i_{1}}(t),\sigma_{i_{2}}(t),
 \cdots, \sigma_{i_{k_{i}}}(t)) \eqno{(1)}$$
for $i = 1, \dots, N$,where $\sigma_{i}  \in Z_{2} \equiv \{0,1\}$ is the binary state and
$B_{i} \in Z_{2}$ is a Boolean function at the $i$th node,
randomly chosen from $2^{2^{k_{i}}}$ Boolean functions,
where the probability to take 1 (or 0) is assumed to be $p$ (or $1-p$)
(TABLE 1).
In this paper we used a case of $p=1/2$ for the numerical calculations.
But the treatment for other cases of $p \ne 1/2$ is
straightforward\cite{Kinoshita06}.
If we fix the set of the randomly chosen Boolean functions
$\{B_{i}, i =1, \cdots, N \}$
in the course of time development,
then this model is called "{\it quenched}" model\cite{Kauffman1}.
On the other hand,
if we change each time the set of the randomly chosen Boolean functions
in the course of time development,
then this model is called "{\it annealed}" model
\cite{DerridaPomeau,DerridaWeis,DerridaStauf,BasParisi1}.\\
%

%
\begin{center}
\begin{tabularx}{83mm}{XXXXXX}  \hline \hline
{\em $\sigma_{i_{1}}$}& {\em $\sigma_{i_{2}}$} & {\em $\cdots$}
& {\em $\sigma_{i_{k_{i-1}}}$} & {\em $\sigma_{i_{k_{i}}}$}&
{\em $B_{i}$}\\  \hline
$0$&$0$&$\cdots$&$0$&$0$&$\sigma_{1}$\\
$0$&$0$&$\cdots$&$0$&$1$&$\sigma_{2}$\\
$0$&$0$&$\cdots$&$1$&$0$&$\sigma_{3}$\\
$0$&$0$&$\cdots$&$1$&$1$&$\sigma_{4}$\\
$\vdots$&$\vdots$&$\vdots$&$\vdots$&$\vdots$&$\vdots$\\
$1$&$1$&$\cdots$&$1$&$1$&$\sigma_{2^{k_{i}}}$\\ \hline\hline
\end{tabularx}\\
\end{center}
TABLE 1.  The relationship between the Boolean functions $B_{i}$
and inputs, $\{\sigma_{i_{1}}, \sigma_{i_{2}}, \cdots, \sigma_{i_{k_{i}}}\}$.
Since there are $k_{i}$ $\sigma_{i}$'s, each of which has $0$ or $1$,
there are $2^{k_{i}}$ ways of inputs.
These provide $2^{k_{i}}$$\sigma_{i}$'s of outputs,
each of which is $0$ or $1$, randomly chosen with
a probability of $p$ or $1-p$.
Hence, there are possibly $2^{2^{k_{i}}}$ Boolean functions at each node.
\\
%

%
We then study the dynamics of the "quenched" RBN.
Since there are $N$ nodes in the random network,
there are $2^{N}$ states
$|\psi\rangle = |\sigma_{1},\sigma_{2}, \dots, \sigma_{N}\rangle$
in the system such as $|\psi_{0}\rangle = |0,0,0, \dots, 0\rangle, \dots,
|\psi_{2^{N}-1}\rangle =|1,1,1, \dots, 1\rangle$.
These states form all vertices of a hypercube in $N$ dimensions.
First, we start to define the initial state $|\psi^{(1)}\rangle$
out of the $2^{N}$ states (for example, say $|\psi^{(1)}\rangle = |0,1, 1, \dots, 0\rangle$).
Second, according to the Boolean functions defined on the network,
the initial state evolves to another state $|\psi^{(2)}\rangle$ in the
$2^{N}$ states.
Hence, the time development can be simply denoted as
$$|\psi^{(t+1)}\rangle = B_{t} |\psi^{(t)}\rangle, \eqno{(2)}$$
where each entry in $|\psi^{(t)}\rangle$ can be developed by Eq.(1)
and $B_{t}$ means a Boolean function chosen at time $t$.
Therefore, if we collect all the $2^{N}$ states as a $2^{N}$-dimensional vector
$|\Psi^{(t)}\rangle = (|0,0,0, \dots, 0\rangle, \dots, |1,1,1, \dots,
1\rangle)^{T}$, then we can get a matrix representation of Eq.(2) as
$$|\tilde{\Psi}^{(t+1)}\rangle = \hat{B}_{t}|{\Psi}^{(t)}\rangle. \eqno{(3)}$$
Here $\hat{B}_{t}$ is a $2^{N}\times 2^{N}$ matrix
that all components are only 0 or 1,
and $|\tilde{\Psi}^{(t+1)}\rangle$ means a state vector
whose components are degenerate
such that the mapping of Eq.(3) is not a one-to-one but a many-to-one
correspondence.
Considering this time developing equation,
we find the cyclic structure of the states such as
the length of the state cycle,
the transient time and the basin size, and so on\cite{KauffmanBook1,Kauffman1,AldanaCK}.
It is obviously difficult to do this procedure by an analytical method in
general.
Therefore, we must do it numerically\cite{AldanaCK}
as well as analytically if possible.
However, there are some important analytical results.
The analytical investigations on the "annealed" models
\cite{DerridaPomeau,DerridaWeis,DerridaStauf,BasParisi1}
showed the existence of a phase transition at
the critical value of $K_{c} = 1/[2p(1-p)]$
\cite{DerridaPomeau,DerridaWeis,DerridaStauf,BasParisi1},
and if we solve it conversely for $p$ then
we obtain the critical value $p_{c} = (1 \pm \sqrt{1-2/K})/2$.
As is described in the introduction,
 the analytical methods for the systems
with special values of the degree of nodes
\cite{DerridaFlyv1,DerridaFlyv2,DerridaFlyv3,DerridaFlyv4,Derrida,Flyvbjerg,
FlyvbjergKjaer,SamuelTroein03,SamuelTroein05}
have been studied in details, already.
%

%
\section{Various network models}

To apply the RBN to that with a given network,
we have to specify what kind of network model we take.
Since there are so many types of networks\cite{AlBarabasi1},
we limit ourselves only to consider
the directed random networks that were first considered by
Kauffman\cite{KauffmanBook1}, the directed scale-free networks, and
the directed exponential-fluctuation networks in this paper.
But the generalization to other networks is straightforward.
In this section, we will give how to generate such directed networks
except the directed random networks, since the generation of those is very well known
\cite{ErdosRenyi1,ErdosRenyi2,ErdosRenyi3,AlBarabasi1,Barabasi04}.
\subsection{Directed scale-free networks with the integer average degree of
nodes}
Let us first consider the directed scale-free networks.
In this case, we adopt a little modified version of
the so-called Barab\'{a}si-Albert model\cite{Barabasi},
since we have to treat the directed scale-free networks with
fractional numbers of the average degree of a node (i.e., vertex),
$\langle k \rangle$, such as $1 \le \langle k \rangle \le 2$.
Denote by $\nu_{i}$ the $i$-th node to which we want to put links
and denote by $\nu_{j}$ one of the surrounding nodes.
Then, the input from the $j$-th node to the $i$-th node
is described by the in-going arrow as $\nu_{i} \leftarrow \nu_{j}$,
while the output from the $i$-th node to the $j$-th node is described by the out-going arrow as
$ \nu_{i} \rightarrow \nu_{j}$.
Denote by $k_{i}^{in}$ ($\equiv k_{j \rightarrow i}$) the input degree of
the $i$-th node.
Denote by $k_{i}^{out}$ ($\equiv k_{j \leftarrow i}$) the output degree of
the $i$-th node.
We note that from simple consideration,
when one input link to the $i$-th node is put between
the $i$-th node and the $j$-th node, it becomes
one output link for the $j$-th node at the same time.
Therefore, if one input link is increased,
so is one output link simultaneously.
Let us consider the case that the average degree of nodes,
$\langle k \rangle$, is an even integer, such as $\langle k \rangle = 2n$.
(A1) We initially start with $m_{0}$ ($\ge n$) nodes for seeds of the system.
We assume that both input and output links
are simultaneously linked between all of the initial seed nodes. Therefore,
the total link numbers for the input and the output are
$n(n-1)/2 \equiv L_{0}$, respectively.
(A2) Next, every time when we add one node to the system,
$m = n$ ($\le m_{0}$) new links are randomly chosen in the previously
existing network,
according to the preferential attachment probability
for the output network,
$$\Pi_{i}(k_{i}^{out})=\frac{k_{i}^{out}}{\sum_{j=1}^{N(t)-1}k_{j}^{out}}.
\eqno{(4)}$$
(A3) Similarly we redo the same procedure for the input network,
according to the preferential attachment probability
for the input network,
$$\Pi_{i}(k_{i}^{in})=\frac{k_{i}^{in}}{\sum_{j=1}^{N(t)-1}k_{j}^{in}}.
\eqno{(5)}$$
We continue the above procedures until
the system size $N$ is achieved.
Therefore, after $t$ steps, we obtain
the total number of nodes as $N(t) = m_{0} + t$ and
the total numbers of links for the input $L^{in}(t)$
and output $L^{out}(t)$ as
$L^{in}(t) = L^{out}(t) = L_{0} + 2nt$, respectively.
Hence, by this we can obtain for the directed scale-free network
$\langle k^{in} \rangle \equiv 2L^{in}(t)/N(t) = 2n$,
$\langle k^{out} \rangle \equiv 2L^{out}(t)/N(t) = 2n$
as $t \rightarrow \infty$.
When $\langle k \rangle = 2n + 1$,
after the procedure (A3) we add one more procedure:
(A4) we redo the procedure (A2) or (A3) with equal probability.
\subsection{Directed scale-free networks with the fractional average degree
of nodes}
Let us next consider the directed scale-free networks with a fractional
average number of degree.
Suppose that $\langle k \rangle$ is fractional such that
$\langle k \rangle = [\langle k \rangle] + (\langle k \rangle)$
where
$[\langle k \rangle]$ means the integer part (say, $n$)
and $(\langle k \rangle)$ the fractional part (say, $f$)
such that $\langle k \rangle = n + f$, where $0 < f < 1$.
In this case,
(B1) we first follow the same procedure (A1) in the previous subsection A.
(B2) Second, we add one node to the system at each step of time.
Every time when a new node is added,
we have to define the node to give input or output.
For this, we randomly choose input or output with equal probability of $0.5$.
(B3) Third, if the chosen case is input (output) for the node,
then we follow the procedure (A2) [(A3)] in the previous subsection A.
Then, we place the input (output) links with equal probability $1$
among the chosen links.
(B4) Fourth, if the not-chosen case is output (input) for the node,
then we follow the procedure (A3) [(A2)] in the previous subsection A.
Then, we place the output (input) links with equal probability
$f/n$ ($0 < f < 1$) among the
chosen links.
(B5) Fifth, go back to (B2) and redo the same procedures
until the system size $N$ is achieved.
Then, after $t$ steps, we obtain
the total number of nodes $N(t) = n + t$,
the total number of input links $L^{in}(t) = L_{0} + (n + f)t$,
and the total number of output links $L^{out}(t) = L_{0} + (n + f)t$,
respectively.
Hence, we can obtain the average input and output degrees of nodes
$\langle k^{in} \rangle \equiv L^{in}(t)/N(t) = n + f$
and $\langle k^{out} \rangle \equiv L^{out}(t)/N(t) = n + f$
as $t \rightarrow \infty$, respectively.
Using the above method, we can construct a scale-free network
with a fractional average degree of nodes $1 < \langle k \rangle < 2$.
For example, consider the case of $\langle k \rangle = 1.4$.
In this case, we just take $n = 1$ and $f = 0.4$.
\subsection{Directed exponential-fluctuation networks}
Let us consider the directed exponential-fluctuation networks.
We can follow the same procedure for both the cases of
the integer and fractional average degrees of nodes,
replacing the probabilities of Eqs.(4) and (5) by
$$\Pi_{i}(k_{i}^{in})=\Pi_{i}(k_{i}^{out}) = \frac{1}{m_{0} + t}.
\eqno{(6)}$$
The exponential-like distributions are often observed in some real-world
networks \cite{sen03}.
The generalization to other networks can be straightforward\cite{wilk05}.
\begin{figure}
\includegraphics[width=8cm]{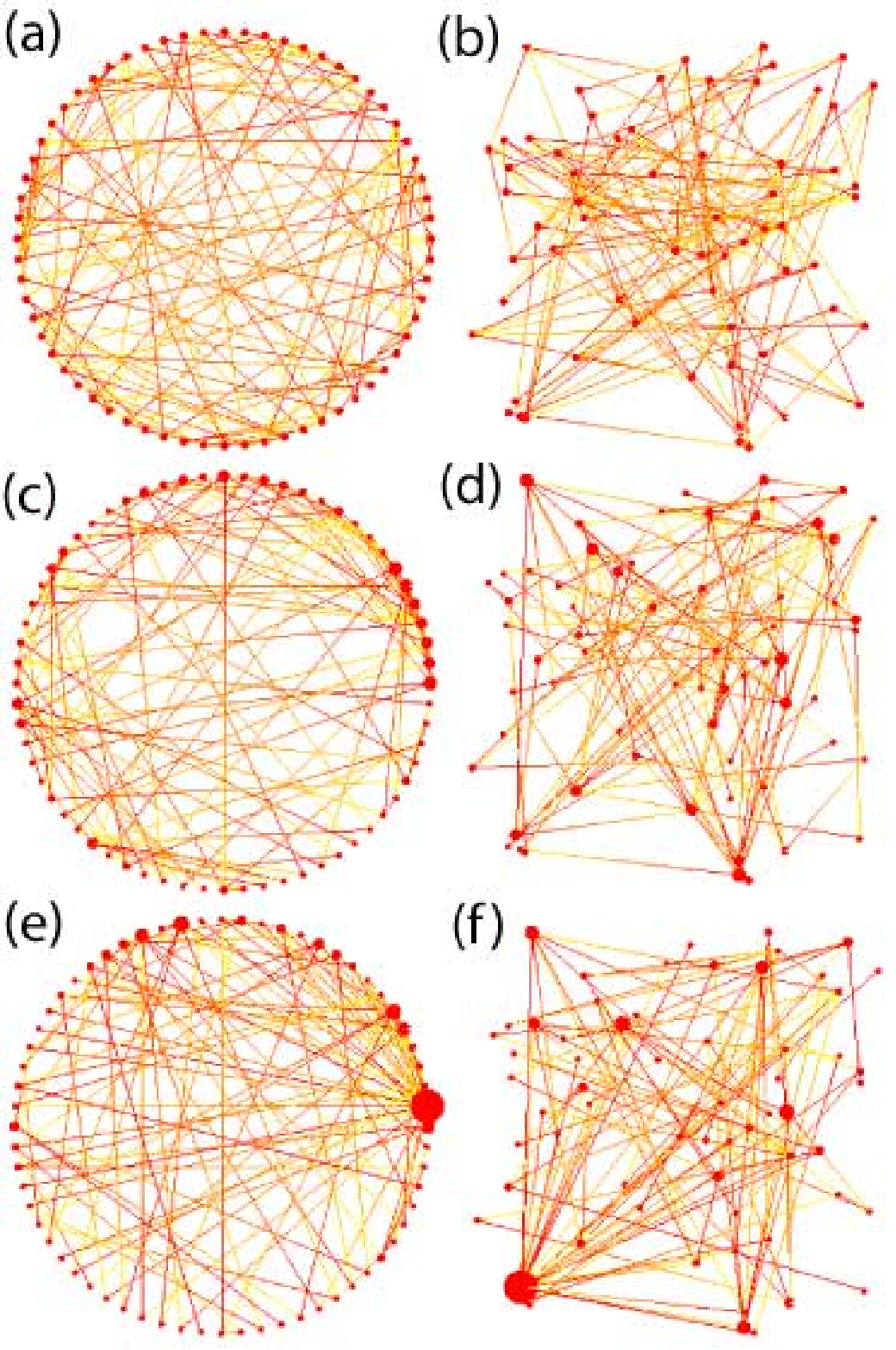}
\caption{
(Color online)
Typical examples of directed networks are shown for
the size of $N = 64$.
(a),(b) Random network with $K = 2$;
(c),(d) Exponential-fluctuation network with $\langle k \rangle = 2$;
(e),(f) Scale-free network with $\langle k \rangle = 2$.
We show same network by two kinds of representation.
For (a), (c) and (e) the nodes are located on the circumference
with equal distance.
For (b), (d) and (f) the nodes are randomly distributed in the square.
Each node is represented as a bold point with size in proportion to the
number of the input links.
We represent input(output)-side of the links with deep(faint) color
such that the direction of a link is denoted by the color gradation
from deep color(output) to faint color(input).
}
\label{fig:1}
\end{figure}
\\
%

%
Fig.1 shows typical examples of the directed networks
where the total number of nodes $N = 64$ is used:
(a),(b) the directed random network with the degree of nodes, $K = 2$;
(c),(d) the directed scale-free network with the average degree
of nodes, $\langle k \rangle = 2$;
and
(e),(f) the directed exponential-fluctuation networks
with the average degree of nodes, $\langle k \rangle = 2$,
respectively.
\section{Cycle distributions}
Now we are going to apply the above-mentioned
"quenched" Kauffman's Boolean dynamics
to the directed random networks,
the directed scale-free networks, and
the directed exponential-fluctuation networks.
The first one provides the famous "quenched" RBN,
the second one the "quenched" SFBN model,
and the third one the "quenched" exponential-fluctuation random boolean network(EFRBN) model\cite{KauffmanBook1}.
Before going to present the numerical results,
let us explain the calculation method for
obtaining the lengths of the state cycles as follows:
(i) {\it Realizations}:
We first prepare $1,000$ sample networks
with the size of $N$ and investigate them in order.
(ii) {\it Initial conditions}:
Second, we randomly choose an initial state $|\Psi^{(0)}\rangle$
out of the $2^{N}$ states.
(iii) {\it The "quenched" Boolean dynamics}:
Third, we calculate the "Boolean dynamics" [given by Eq.(2)] repeatedly
from $10$ times to $100,000$ times,
where we fix a special set of the Boolean functions on the network.
This means that we treat the "quenched" model.
(iv) {\it State cycles}:
Fourth, we investigate whether there exist the states that
belong to the equivalent state in the data.
Since the "Boolean dynamics" is deterministic,
if the state $|\Psi^{(t_1)}\rangle$ at time $t = t_1$ is the same as
the state $|\Psi^{(t_1 + s)}\rangle$ at $t = t_1 + s$,
then the state $|\Psi^{(t_1 +1)}\rangle$ at $t = t_1 + 1$ becomes the same
state as the state $|\Psi^{(t_1 + s +1)}\rangle$ at time $t=t_1 + s + 1$.
Therefore, the length of the state cycles satisfying this condition is $s + 1$.
Such calculations are performed for all network samples.
\subsection{Histograms of the lengths $\ell_c$ of the state cycles}
\begin{figure}
\includegraphics[width=8cm]{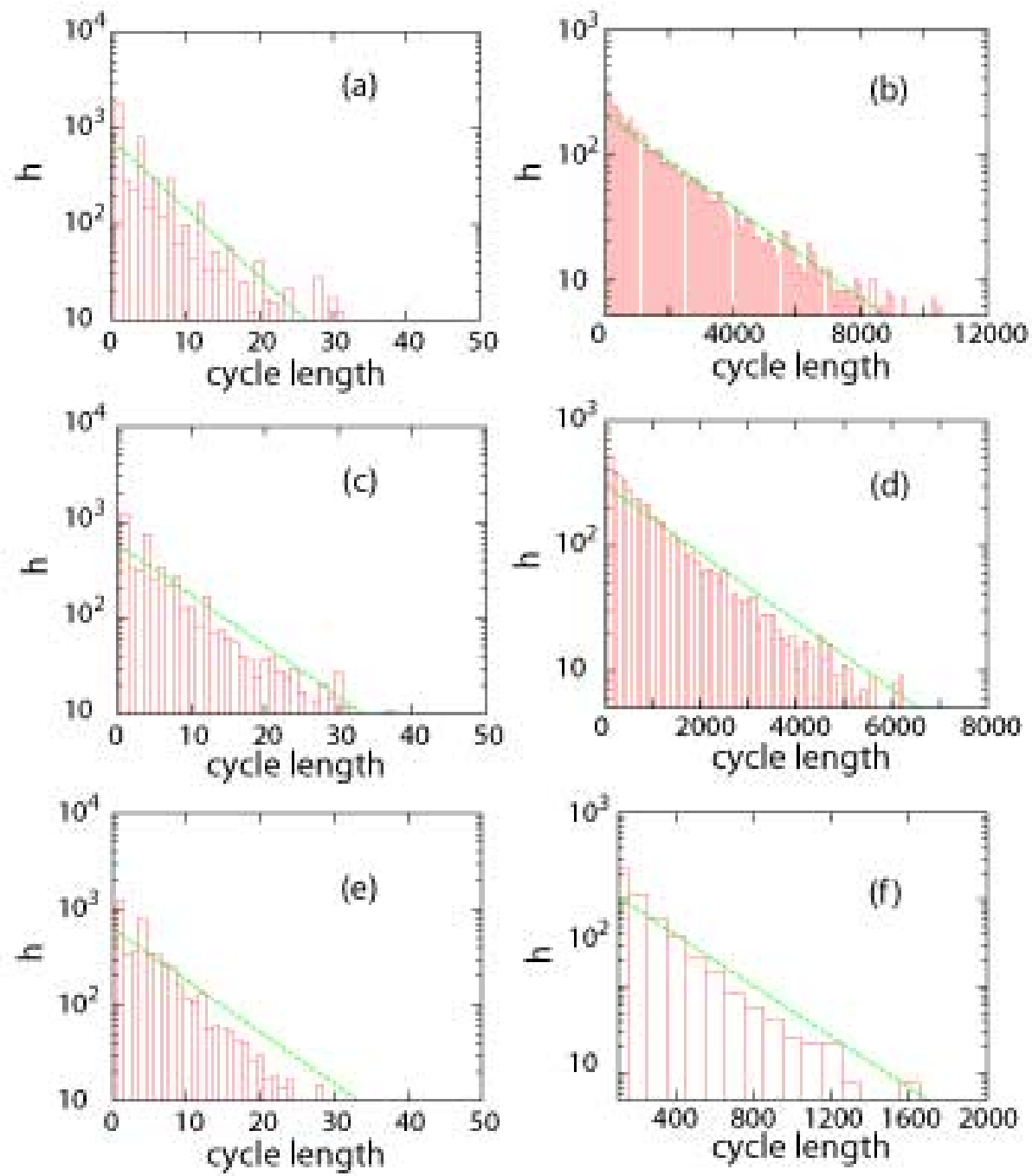}
\caption{
(Color online)
Histograms $h$ of the lengths $\ell_c$ of state cycles
in various types of the directed networks. The network size is $N=40$.
Each histogram is generated by $10^3$ different sets of the Boolean functions
and five different network structures.
The maximum iteration number of the Boolean dynamics is $10^5$
until the convergence to the cycle is realized.
(a)-(b) RBN with $K = 2, 4$;
(c)-(d) EFRBN with $\langle k \rangle =
2, 4$;
(e)-(f) SFRBN with $\langle k \rangle = 2, 4$.
Here in (a)-(b) $2^{2^{K}}$ Boolean functions of $K$ variables are used
equiprobably,
while in (c)-(f)
$2^{2^{k_{i}}}$ Boolean functions of $k_{i}$ variables are used equiprobably
at the $i$-th node.
The lines are the exponential fittings to the numerical data.
The mesh sizes of the histograms are 1 for $K = 2$ and $\langle k \rangle =
2$,
$10^2$ for $K = 4$ and $\langle k \rangle = 4$.
}
\label{fig:2}
\end{figure}
Fig.2 shows the histograms of the lengths $\ell_c$ of the state cycles
in  the original "quenched" RBN
with $K = 2,4$ and in the "quenched" SFRBN and EFRBN where the average degree of nodes $\langle k \rangle = 2, 4$.
We can see that the functional forms of the distributions
seem exponential type.
The tail of the distribution depends on the fluctuation property of the degree of the nodes.
We found that in comparing among the three types of the network structures
the maximum length of the state cycle becomes longer
as the fluctuation in the degrees of nodes becomes larger.
We try to investigate the more accurate functional form of the distribution
and
the transition depending on the degree of nodes $K$ and $\langle k \rangle$.
\subsection{The median $\bar{m}$, the mean value $\langle \ell_c \rangle$,
and the standard deviation $\sigma$
of the distributions of the state cycle lengths}
\begin{figure}
\includegraphics[width=8cm]{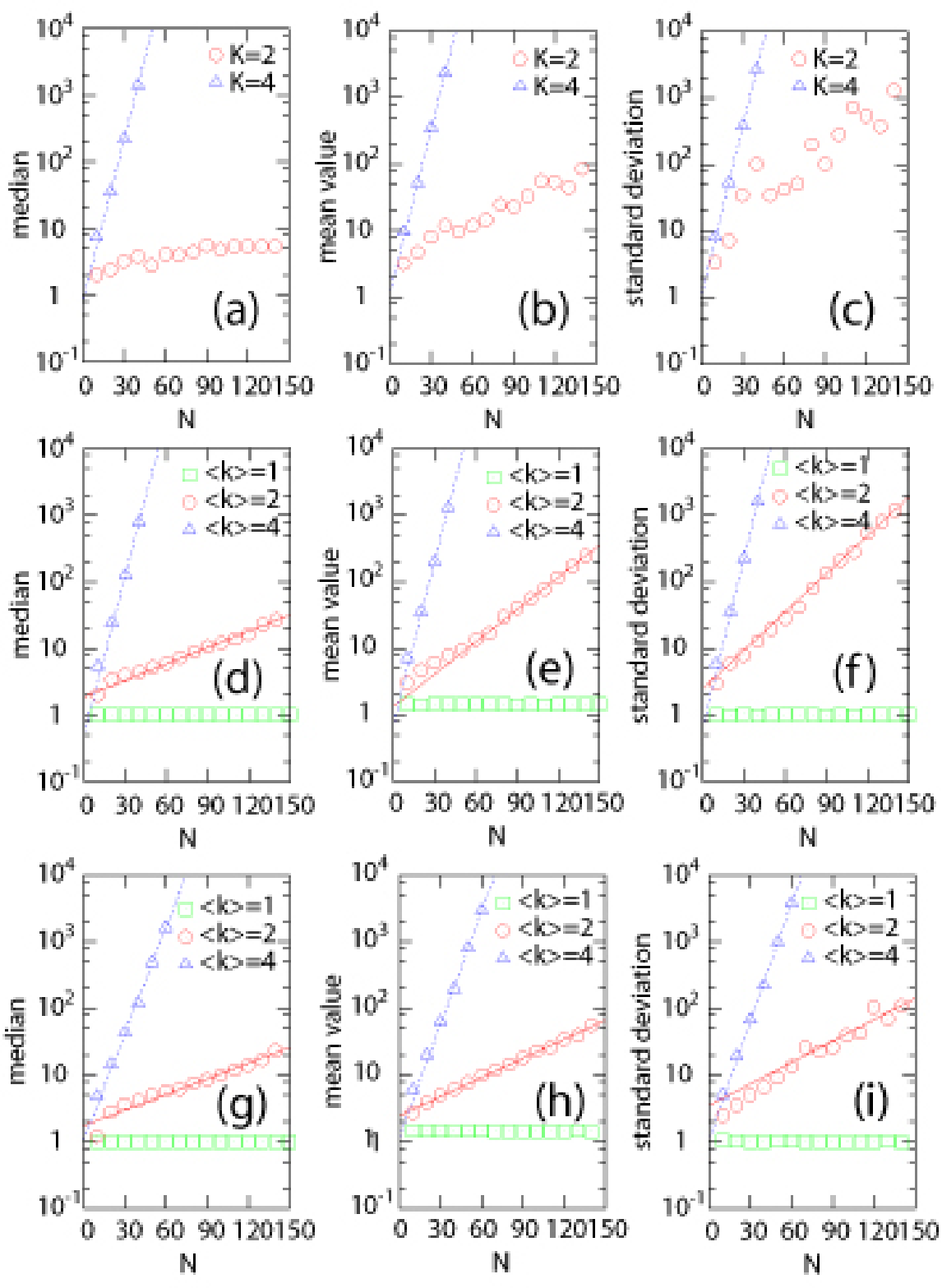}
\caption{
(Color online)
Semilog-plots of the
the median $\bar{m}$, the mean value $\langle \ell_c \rangle$,
and the standard deviation $\sigma$
of 5000 samples of the state cycles with respect to
the total number $N$ of nodes for the various directed networks,
respectively.
(a)-(c) are shown for the RBN
with $K = 2, 4$.
(e)-(f) are shown for the EFRBN
with $\langle k \rangle =1, 2, 4$.
And (g)-(i) are shown for the SFRBN
with $\langle k \rangle =1, 2, 4$.
The lines are best exponential fittings,
$\bar{m} \sim e^{aN}$, $\langle \ell_c \rangle \sim e^{bN}$, $\sigma \sim
e^{cN}$,
to the numerical data
except for the RBN with $K = 2$.
The best fitted exponents are $a=0.02, b=0.037, c=0.044$ for the EFRBN with $\langle k\rangle =2$;
$a=0.0168, b=0.022, c=0.025$ for the SFRBN with $\langle k\rangle =2$;
$a=0.163, b=0.193, c=0.194$ for the RBN with $K=4$;
$a=0.170, b=0.179, c=0.194$ for the EFRBN with $\langle k\rangle =4$;
$a=0.123, b=0.131, c=0.136$ for the SFRBN with
$\langle k\rangle =4$.
}
\label{fig:3}
\end{figure}
%

%
Fig.3 shows the median $\bar{m}$, the mean value $\langle \ell_c \rangle$,
and the standard deviation $\sigma$
of the distributions of the state cycle lengths with respect to the total number $N(\sim 150)$ 
of nodes for the various directed networks, respectively.
Fig. 3(a) shows that in the RBN with $K=2$ the median grows
as $\sqrt N$ in relatively small $N$ region and it grows as proportional to $N$ in the large $N$.
Whether or not the behavior of $\sqrt N$ is valid is very delicate
since we have to always stem this from the numerical data of the finite size systems.
The more details will be shown elsewhere\cite{Kinoshita06}.
On the other hand,  in the RBN with $K=4$ the
median grows exponentially with $N$ as $\bar{m} \sim e^{aN}$.
We have observed that as $K$ increases, the growth type of the median exhibits a transition 
from algebraic type to exponential one in the RBN.
In the EFRBN and the SFRBN, the median grows exponentially with $N$ for $\langle k\rangle =2$ and $4$,
and such a transition can not be observed
in the range between $\langle k\rangle =2$ and $\langle k\rangle =4$.
Therefore, we conjecture that based on Fig.3,
the transition takes place in the range
between $\langle k\rangle =1$ and  $\langle k\rangle =2$
in the RBN and the EFRBN.Note that $N$ dependence for $\langle k \rangle =1$ can be analytically derived.
(See appendix A.)
Furthermore, we can see the mean value $\langle \ell_c \rangle$ and
the standard deviation $\sigma$ grow exponentially
with $N$ such as
$\langle \ell_c \rangle \sim e^{bN}$, $\sigma \sim e^{cN}$,
in all cases.
It is difficult to distinguish clearly the dynamical properties between
the SFRBN and the EFRBN
when the network size $N$ is as small as $N \sim 150$.
\subsection{The relationship between the mean value $\langle \ell_c \rangle$
and the standard deviation $\sigma$}
\begin{figure}
\includegraphics[width=8cm]{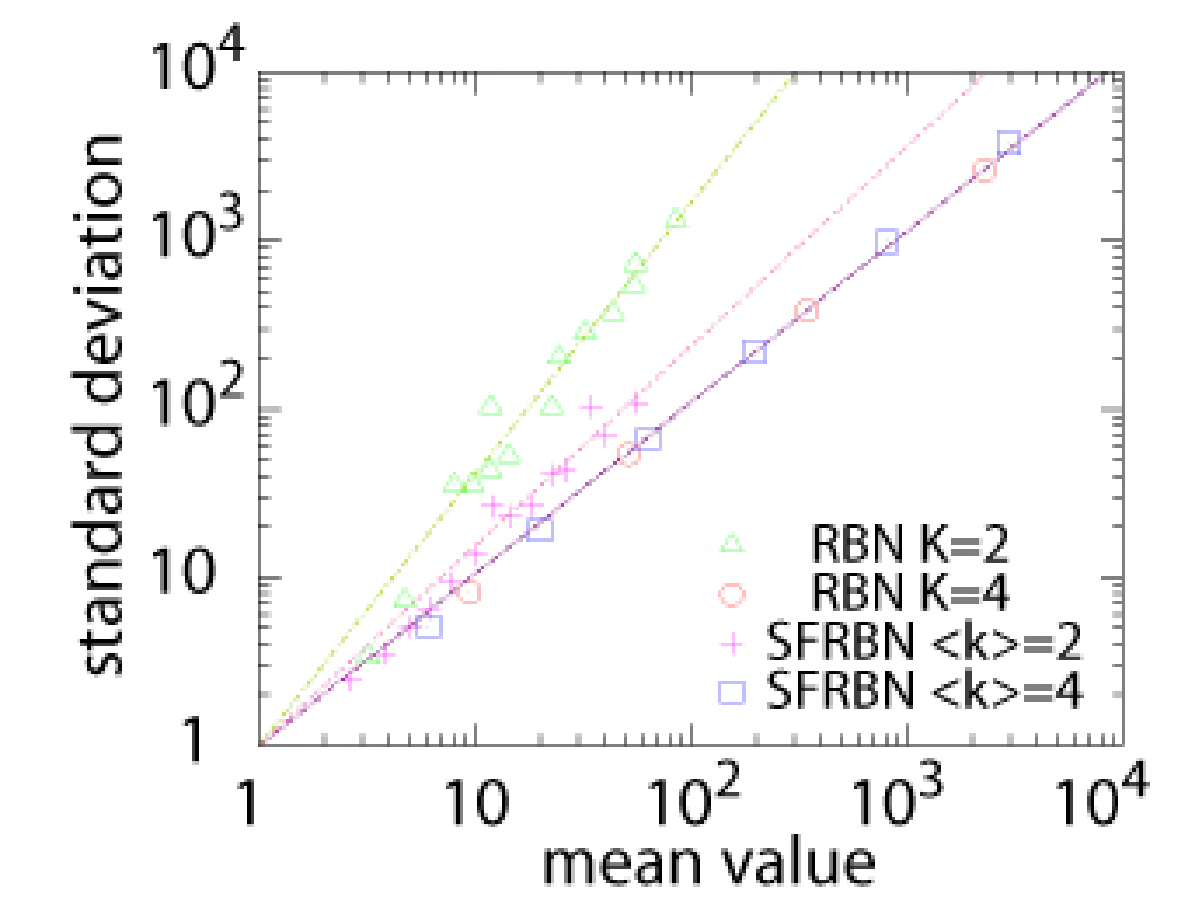}
\caption{
(Color online)
The relationship between the mean value $\langle\ell_c\rangle$
and the standard deviation $\sigma$
of the distributions of the state cycle lengths with respect to
the total number $N$ of nodes
for the RBN with $K = 2,4$
and the SFRBN $\langle k \rangle = 2, 4$, respectively.
The lines are fitting to the numerical data as $\sigma \sim \langle\ell_c\rangle^\beta$.
The best fitted exponents are $\beta=1.61,1.02$
for the RBN with $K=2,4$,
and  $\beta=1.19,1.02$
for the SFRBN with $\langle k\rangle =2,4$, respectively.
The result for the EFRBN is almost the
same as that for the SFRBN.
}
\label{fig:4}
\end{figure}
%

%
Fig.4 shows the relationship between
the mean value $\langle \ell_c \rangle$
and the standard deviation $\sigma$
of the distributions of the lengths of the state cycles with respect to
the total number $N$ of nodes for
the RBN
and the SFRBN, respectively.
We have found that the relationship is fitted by
the following expression:
$\sigma = \langle \ell_c \rangle^{\beta}$.
The best fitted exponent $\beta$ is nearly equal to unity
for $K=4$ and $\langle k\rangle =4$.
As a matter of fact,
we can guess that
in the continuum limit of the relatively large $N$,
the distribution $P(\ell_c)$ approaches the exponential form
 $P(\ell_c) = \alpha \exp(-\alpha \ell_c)$, as $K$ and $\langle k \rangle$
increases.
In the exponential distribution, the mean value $\langle\ell_c \rangle$
and the standard deviation $\sigma$ can be represented in terms of
the single parameter $\alpha$
such as $\langle\ell_c \rangle = \sigma = 1/\alpha$.
\subsection{Plots of the median value $\bar{m}$}
\begin{figure}
\includegraphics[width=8cm]{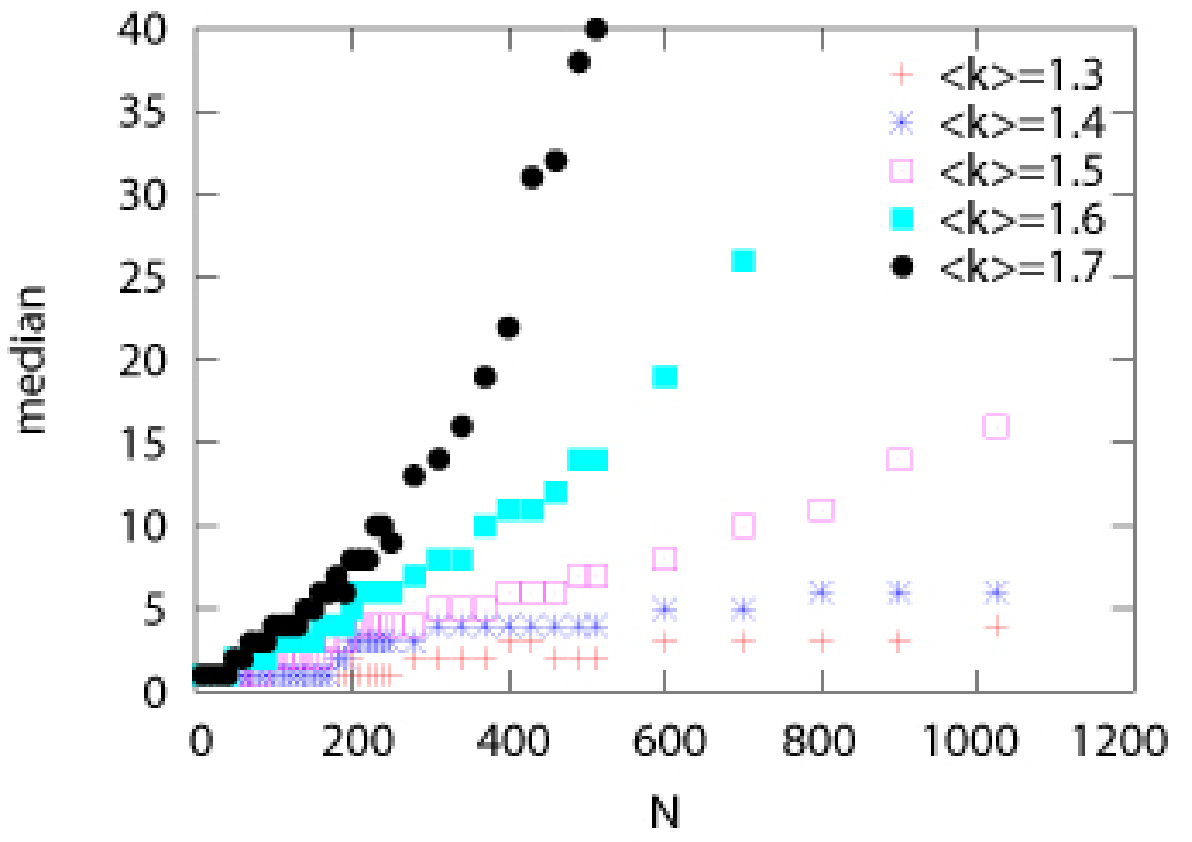}
\caption{
(Color online)
Plots of the median value $\bar{m}$ of 1000 samples of
the lengths of the state cycles
with respect to the total number $N$ of nodes for the directed scale-free
networks
with $\langle k\rangle = $1.3,1.4,1.5,1.6 and 1.7.
The results for $N$ up to 1024 are shown.
We used a cut-off attractor cycle length $\ell_{cut}=10^5$
due to the numerical difficulty to detect the large cycle length.
For the 10-20 network samples, we did not detect the attractor cycle lengths.
However, it seriously does not influence on the median of the distribution.
}
\label{fig:5}
\end{figure}
%

%
As shown in Fig.4 it is clear that the standard deviation $\sigma$ is larger
than the mean value $\langle\ell_c \rangle$ for $K=2$ and $\langle k\rangle = 2$.
In the cases, the numerical mean values are sometimes without
credibility due to the lack of the samples with large cycle lengths.
Such an imperfection of the self-averaging is
well-observed for the distributions with a long-tail.
Therefore, we investigate the relatively stable median
as representative values to characterize the distributions of the attractor
lengths, instead of the mean,  for the large networks.


Fig.5 shows the median value $\bar{m}$ of the distribution of state cycle
lengths with respect to the total number $N$ of nodes for the relatively large
SFRBN ($N \sim 10^3$).
We found that there is a transition in a range of $1.4 < \langle k\rangle_c
< 1.7$,  dividing the
$N^\alpha (\alpha <1)$ growth and $N^\alpha (\alpha >1)$ growth in
polynomial growth in the ordered phase.
Note that some biological data suggest slow growth as $N^\alpha (\alpha \leq
1)$ in the relatively
small network size ($N \sim 10^3$).

It is well-known that the scale-free topologies are ubiquitously in nature
and the degree exponents lie in between 2 and 3.
Furthermore, one of the important property in the scale-free topology is the
existence of the highly connected hub as seen in the yeast synthetic network and so on \cite{Tang04}.
The realistic systems do not contain enough nodes to closely approximate
the true transition, therefore, the finite-size behavior is important.
In fact some biological realistic data suggest that the number of cell types
in organism is crudely proportional to the linear or square-root function
of the DNA content per cell, i.e. $\bar{m} \propto N^\alpha (\alpha \leq 1)$
\cite{KauffmanBook1}, for the finite-size $N \sim 10^3$.

\begin{figure}
\includegraphics[width=8cm]{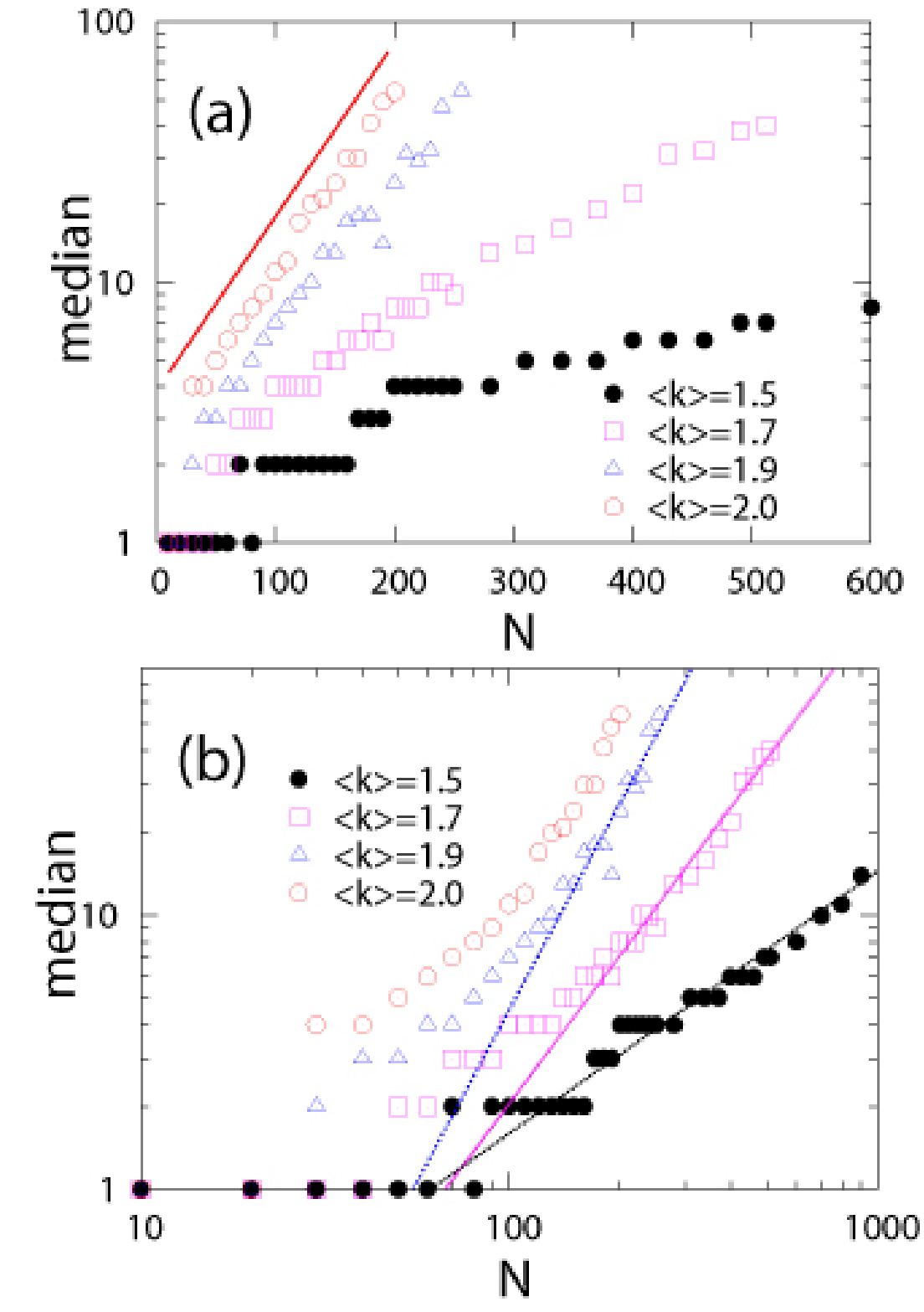}
\caption{
(Color online)
(a)Semilog-plots and (b)log-log-plots of the median $\bar{m}$ with
respect to the total number of nodes for some values $\langle k\rangle$ near $\langle k\rangle =2$ for the scale-free network in Fig.5.
Apparently, the case with $\langle k\rangle =2$ exponentially increases with respect to $N$.
The straight line in (a) shows the exponential growth for a guide for eyes,
 and the straight lines in (b) show the polynomial growth ($\bar{m} \propto N^\alpha$) with the slopes $\alpha =2.5, 1.8, 0.98$
from top to bottom.
}
\label{fig:6}
\end{figure}

Finally, we investigate a transition between the polynomial growth in the
ordered phase and the exponential one in the chaotic phase.
Figure 6 shows the semi-log and log-log plots of the data near $\langle
k\rangle_2 \sim 2$ in Fig.5.
It is suggested that the function form $\bar{m}(N)$ has an polynomial form for
$\langle k\rangle <2 $ and approaches the exponential form as $\langle
k\rangle \to 2$
in the quenched SFRBN.
The results are consistent with the critical value $K_c=2$ given
for annealed SFRBN in other reports \cite{LuqueSole,Aldana1,AldanaCl,Aldana2}.
In this section we focused on the scaling of attractor length near critical
value $\langle k \rangle \leqq 2$. The other quantities are also useful measure for the
transition of the Boolean dynamics. (See appendix B.)
%

%
\section{Conclusions}
In conclusion, we have studied the Boolean dynamics of the "quenched"
Kauffman model with
directed scale-free networks(SFRBN), comparing with ones of
the directed random networks(RBN)
and the directed exponential-fluctuation networks(EFRBN).
We have numerically calculated the
distributions of the lengths of the cycles and its changes
as the network size $N$ and the average degree of the node increase.
We have found that the median, the mean value and the standard deviation
grows exponentially with $N$
in the EFRBN and
the SFRBN with $\langle k\rangle =2,4$,
where the function forms of the distributions are almost exponential.

>From our results we conclude that a transition occurs near $ \langle k \rangle \sim 2$ in the SFRBN.
The result is consistent with that in the "annealed" SFRBN.
%

%
In this paper we dealt with the directed graphs that correspond to
the asymmetric adjacency matrices in the network theory.
However, the quite different distributions of the state cycle lengths
are observed in the undirected random networks.
Therefore, it is also interesting to study
the properties of the distributions of the state cycle lengths in
the undirected scale-free networks as well.
The details will be given elsewhere \cite{Kinoshita06}.
We have numerically investigated the finite-size networks ($N \sim 140-1024$).
In fact, it is very difficult to study the indefinitely large size of systems.
This seems the disadvantage of our approach of the finite size systems.
However, practically speaking,
there are a lot of scale-free-type biological networks
with size $N \sim O(10^2) - O(10^3)$, such as
the metabolic reaction networks of the bacterium E.coli and of
the Yeast protein interaction networks, etc\cite{Tang04,Barabasi04}.
Thus, we believe that our results stemmed from the finite size
systems might play an important role to study such real network
systems in Nature. As we mentioned in the introduction, the asynchronous
updating rules are more plausible for biological phenomena.
It is interesting to expand the investigation to the asynchronous version.
\acknowledgments
We would like to thank Dr. Jun Hidaka for collecting many relevant papers
on the Kauffman model and the related topics.
One of us (K. I.) would like to thank Kazuko Iguchi for her continuous
financial support and encouragement.
%

%
\appendix

\section{A case with $\langle k \rangle =1$}
In this appendix, we compare the numerical results in SFRBN with $\langle k \rangle =1$
with analytically result in RBN with $K=1$.

In Fig.7, we show the finite-size effect of the mean value $\langle \ell_c \rangle$.
The analytical result $\langle \ell_c \rangle \simeq 1.44$ in RBN with $K=1$
have been derived based on the probability for distribution of
information conserving loops by Flyvbjerg and Kjaer \cite{FlyvbjergKjaer}.
It is found that the numerical results converge a value $1.44$
as the system size increases in both SFRBN and exponential-fluctuation
network.

\begin{figure}
\includegraphics[width=8cm]{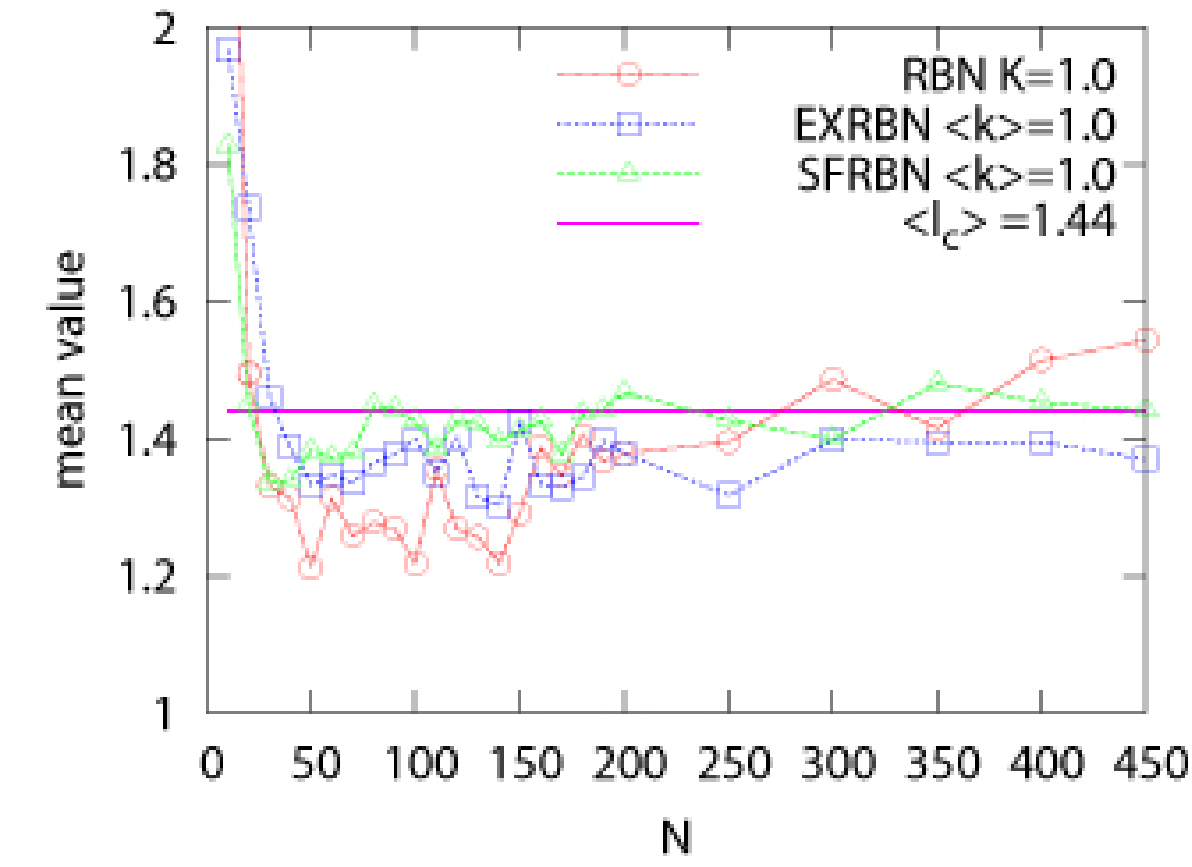}
\caption{
(Color online)
Mean value $\langle \ell_c \rangle$ of attractor lengths in RBN and SFRBN with
$\langle k \rangle =1$. The analytical result $\langle \ell_c \rangle = -\rho/(1 - \rho)\log(1 - \rho)\simeq 1.44$ 
for $\rho = 1/2$ in the annealed RBN is also overplotted.
}
\label{fig:7}
\end{figure}

As mentioned in Sect.IV, we constructed the distribution of the
attracter length by different networks with $\langle k \rangle =1$, i.e. case (A) in
introduction.Then the states converge into a point attractor
with period unity, $\ell_c=1$, through the transient states
in a lot of the network samples.We can partially see the result in the distribution in \ref{fig:2}.
Accordingly the median $\bar{m}$ over the distribution
is always unity independent of
syatem size $N$ in our case.

\section{Derrida plots}
In this appendix, we give the numerical results of Derrida plot
in the SFRBN.
The Derrida plot, analogous to the Lyapunov exponent in the continuous
dynamics,
measures the divergence of trajectory based on the normalized Hamming
distance
between two distinct states $\sigma_i^{1}(t),\sigma_i^{2}(t), i=1,2,..,N$,
\begin{equation}
H(t) = \frac{1}{N}\sum_i^N |\sigma_i^{1}(t)-\sigma_i^{2}(t)|.
\end{equation}
For a sample of the random initial states pairs, the average $H(t+1)$ is
plotted against $H(t)$, and the plot is
repeated for increasing $H(t)$.
A curve above the main diagonal $H(t+1)=H(t)$
indicates a divergent trajectory and chaotic.
The area below the diagonal $H(t+1)=H(t)$ is called the ordered region,
because the trajectories converge in the state space.
A curve tangential to the main diagonal $H(t+1)=H(t)$
indicates a critical dynamics.
%

%
\begin{figure}
\includegraphics[width=8cm]{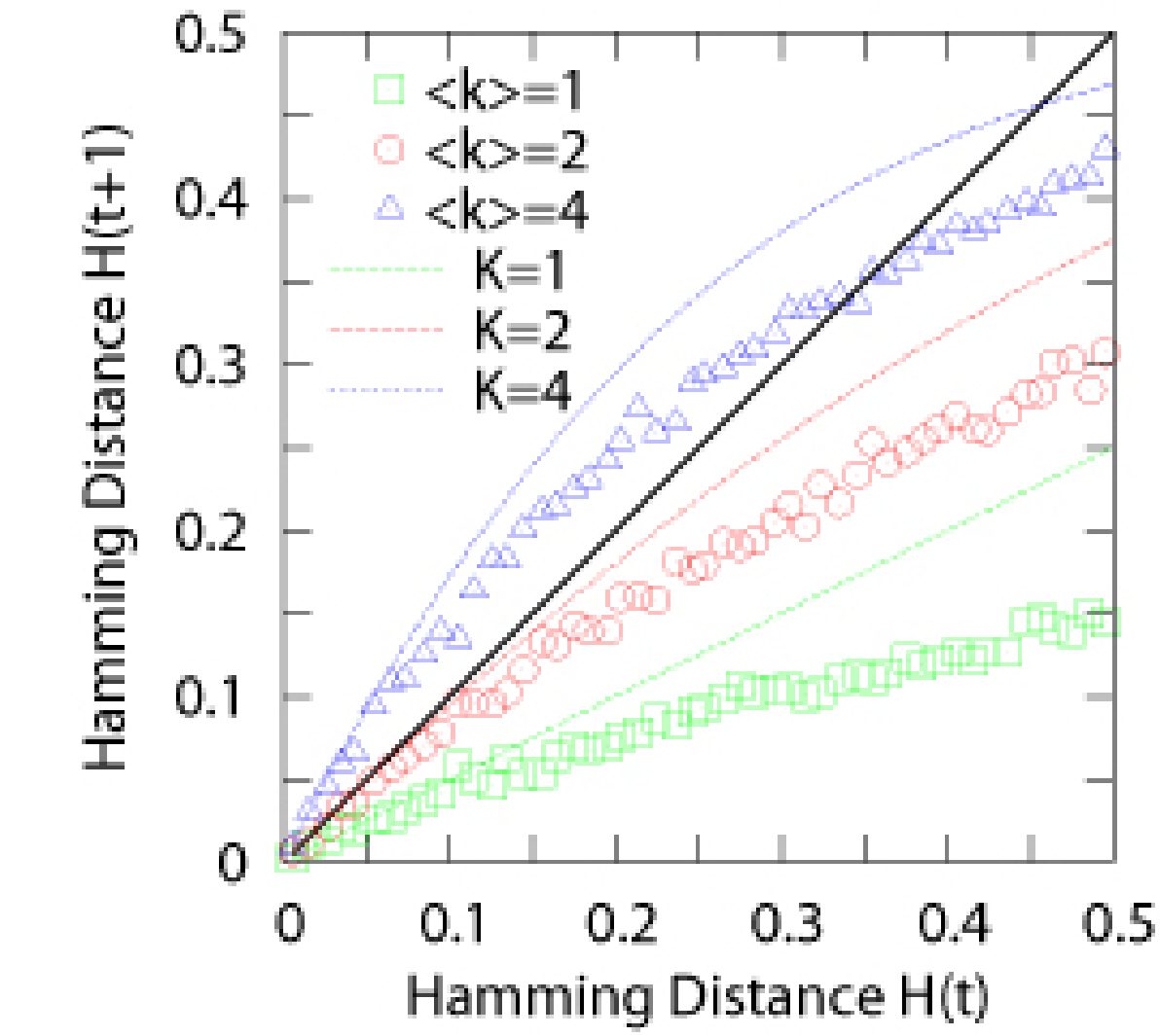}
\caption{
(Color online)
Derrida plots of the SFRBN with
$\langle k \rangle =1$, $\langle k \rangle =2$ and $\langle k \rangle =4$.
The analytical curves for the RBN with $K=1$, $K=2$ and $K=4$ are
also overplotted.@A line $H(t+1)=H(t)$ is the dividing line
between order and chaos. It is clear that $K = 2$ lies directly on this
line,
the system size is $N=1024$ and the number of the initial
states for averaging is 2000.
}
\label{fig:8}
\end{figure}
%

%
Figure 8 shows the Derrida plots corresponding to the
SFRBN with $\langle k \rangle =1,2,4$. It is apparent that the SFRBN with
large $\langle k \rangle$ exhibits more chaotic behavior than that with small $\langle k \rangle$.
It is important
to note that the SFRBN is more ordered than the RBN compared with the cases
with $K=\langle k \rangle$.
The Derrida coefficient is defined by $D_c=\log s$, where $s$ is the slope
of the Derrida plot (curve) at the origin. If $s = 1$
then $D_c = 0$, giving the critical evolution.
The slope of the Derrida plot for the SFRBN with $\langle k \rangle =1$ is approximately
unity. The result is consistent with the occurence of the
transition at $\langle k \rangle =2$ in the function form of $\bar{m}(N)$
observed in Subsec. IV.D.
The other quantities such as,
the density of frozen nodes and the robustness against perturbation,
are also important
indicators for the dynamical behavior in the RBN and the SFRBN.

\end{document}